\documentclass[bibyear]{aa}
\usepackage[varg]{txfonts}
\usepackage{graphicx}
\def\I{\mbox{IRAS\,20126+4104}}

\def\Msun{\mbox{$M_\odot$}}
\def\Rsun{\mbox{$R_\odot$}}
\def\Lsun{\mbox{$L_\odot$}}
\def\WAT{H$_2$O}

\def\METH{\mbox{CH$_3$OH}}

\def\HM{\mbox{H$_2$}}
\def\HII{H{\sc ii}}

\def\kms{\mbox{km~s$^{-1}$}}

\def\mic{\mbox{$\mu$m}}

\def\d{{\rm d}}

\begin{document}
\title{
Evidence for a bloated massive protostar in \I
}
\author{
        R.~Cesaroni 
}
\institute{
 INAF, Osservatorio Astrofisico di Arcetri, Largo E. Fermi 5, I-50125 Firenze, Italy
	   \email{riccardo.cesaroni@inaf.it}
}
\offprints{R. Cesaroni, \email{riccardo.cesaroni@inaf.it}}
\date{Received date / Accepted date}

\abstract{
Variability is a well known phenomenon in low-mass young stellar objects,
but in recent years the monitoring of methanol masers and infrared continuum
emission has permitted the detection of both burst-like episodes and
periodic variations also in high-mass (proto)stars. Multi-epoch studies on
large samples of these objects have become possible thanks to the NEOWISE
database, which surveyed the sky in the mid-IR for about a decade.
}{
Our goal is to analyse the mid-IR emission from the well studied massive
protostar \I\ and confirm the hypothesis that such emission is periodic,
as proposed in previous studies.
}{
We take advantage of the NEOWISE, ALLWISE, and Spitzer databases to
obtain 24 images of the 3.4~\mic\ emission from \I\ spanning 19 years,
with $\sim$6-months sampling over a decade. With these data we create a
light curve for each lobe of the bipolar nebulosity/outflow associated
with the protostar.
}{
Our results confirm that the IR emission from \I\ varies regularly with a
period of $\sim$6.8~yr. The period is the same for both lobes, but their
emissions are anticorrelated with a phase difference of $\sim$2.5~yr. The
variation is consistent with that found in previous studies for the 6~GHz
\METH\ masers and the near-IR emission from the lobes.
}{
After discussing four possible ``clocks'' that could determine the observed
periodicity, we rule out all but a model involving rotation of the star
with a spot obscuring $\sim$20\% of the stellar surface. The long rotation
period implies that the 12~\Msun\ protostar is bloated, with a radius of
$\sim$200~\Rsun.
}

\maketitle

\section{Introduction}
\label{sint}

The existence of variable sources in high-mass star forming regions has
been established in the past through the detection of regular and irregular
changes of the emission at radio and infra-red wavelengths. Monitoring of
the water maser emission revealed increases of the intensity by orders
of magnitude on time scales as short as a few days (see e.g. Felli
et al.~\cite{felli07} and references therein). Although more stable,
class~II methanol masers were also found to have explosive behaviour in a handful
of objects (Fujisawa et al.~\cite{fuji15}; Hunter et al.~\cite{hunt17};
Sugiyama et al.~\cite{sugi19}; Hirota et al.~\cite{hiro22}), but more than
30 sources are known to undergo regular variations with periods ranging
from $\sim$24~days to $\sim$4.4~yr (see Tanabe \& Yonekura~\cite{tayo24}
and references therein). Sometimes variations of the IR intensity have been
observed in association with the increase of the \METH\ maser flux density
(Stecklum~\cite{steck16}; Caratti o Garatti et al.~\cite{cagana}; Harajiri
et al.~\cite{hara26}) as expected if the inversion of the level populations
is due to radiative pumping.

Recently, the Near-Earth Object Wide-field Infrared Survey Explorer
(NEOWISE) mission has provided the astronomical community with regular
monitoring of the sky at 3.4~\mic\ and 4.6~\mic\ over a decade. These data
triggered a number of searches for variable young stellar objects (YSOs)
that established that variability is quite common over time scales from days
to decades (Park et al.~\cite{park21};
Kulkarni et al.~\cite{kulk26}),
being present in a fraction of
the targets ranging from $\sim$26\% (Neha \& Sharma~\cite{neha25}) to $\sim$32\%
(Lu et al.~\cite{lu24}), depending on the selection criteria.

Besides these statistical studies on large samples of targets, it is also
very instructive to focus on specific objects and use the outcome of NEOWISE
to constrain their physical properties and, possibly, understand their
nature. A good target for this type of study is the protostar \I,
a well studied high-mass YSO driving a precessing jet+outflow and surrounded
by a Keplerian disk
(see Cesaroni et al.~\cite{cesa25} and references therein). The mass of the
protostar is estimated to be $\sim$12~\Msun\ (Chen et al.~\cite{chen}),
consistent with a bolometric luminosity of $\sim$$10^4$~\Lsun\ (Johnston
et al.~\cite{johns}; Cesaroni et al.~\cite{cesa23}). The precession
of the bipolar jet and outflow hints at the presence of a companion
(Shepherd et al.~\cite{shep}; Cesaroni et al.~\cite{cesa05}; Caratti
o Garatti et al.~\cite{caga08}), although much less massive than the
primary (Cesaroni et al.~\cite{cesa23}). Estimates of the distance to \I,
measured with the parallax of \WAT\ masers, range from
1.33$^{+0.19}_{-0.15}$ (Hirota et al.~\cite{vera}) to 1.64$\pm$0.05~kpc
(Moscadelli et al.~\cite{mosca11}). For the present study we adopt the
value with the smallest error, i.e. 1.64~kpc.

\begin{figure}
\centering
\resizebox{8.0cm}{!}{\includegraphics[angle=0]{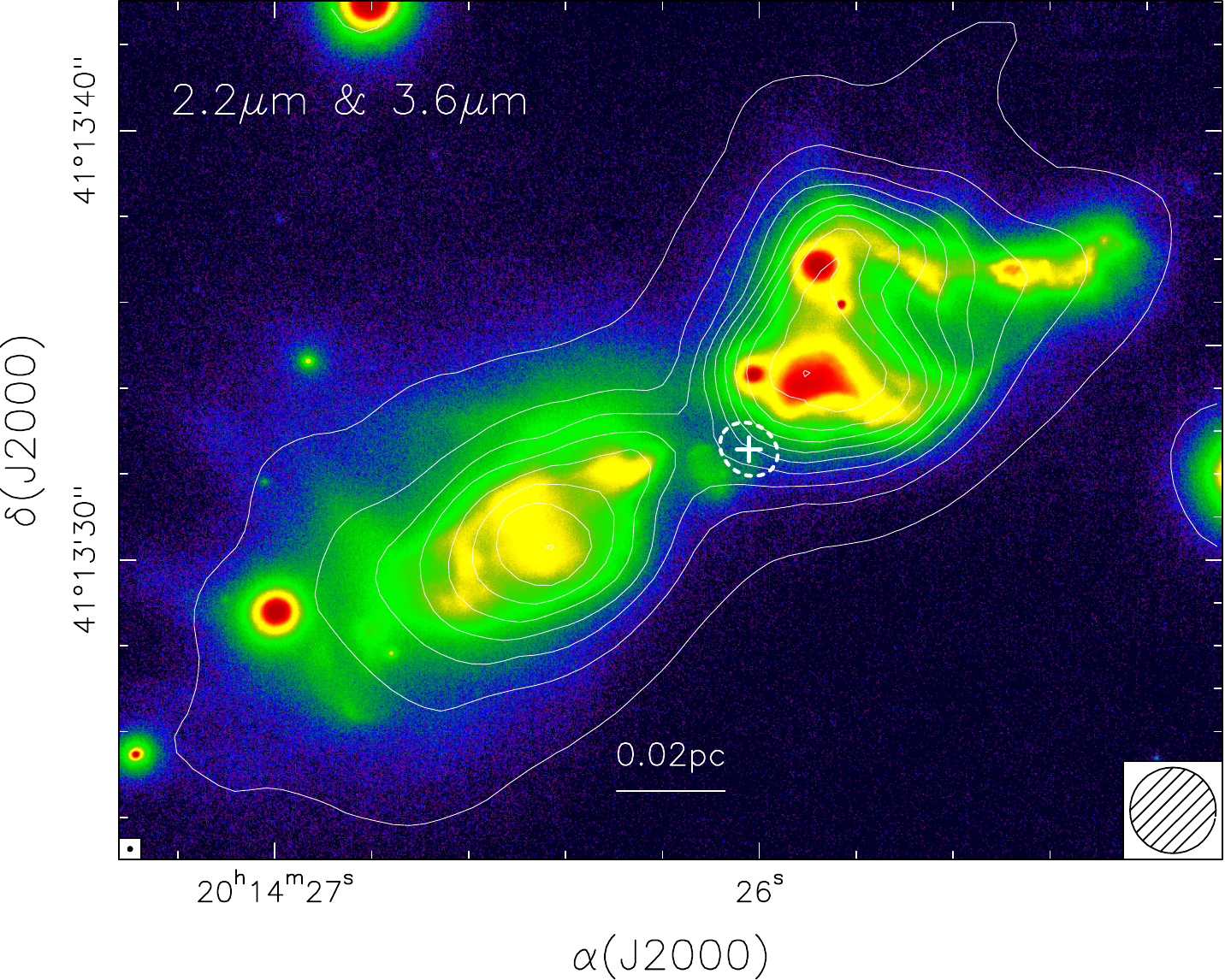}}
\caption{
Overlay of the Spitzer/IRAC image at 3.6~\mic\ (contours) on the LBT/SOUL image
at 2.2~\mic\ from Massi et al.~(\cite{massi23}). Contour levels range from
30 to 430 in steps of 50~MJy/sterad. The dashed ellipse denotes the full width
at half power of the 1.4~mm continuum emission mapped with ALMA by Cesaroni
et al.~(\cite{cesa25}). The cross marks the expected position of the protostar.
}
\label{fjet}
\end{figure}

The main features of the source are shown in Fig.~\ref{fjet}. One sees
that the IR emission traces a bipolar nebula oriented SE--NW, consistent
with the bipolar outflow, and emerging from a dusty core revealed by the
millimeter continuum emission. The protostar is embedded in the core and
lies at the center of a Keplerian disk whose plane is almost perpendicular
to the outflow axis.

A recent study by Szymczak et al.~(\cite{szy24}) reveals a periodic
variation of the Class~II \METH\ maser emission at 6~GHz, with a significant
correlation with the IR emission of the NEOWISE database. If variability
is a clue of the physical mechanisms at work, periodic variability is even
more so. Association with a rotating circumstellar disk and a precessing
jet makes the plot more intriguing and led us to perform a deeper analysis
of the IR emission from \I\ at different epochs. We
describe the archival data used for our study in Sect.~\ref{sarc}, show
the results obtained in Sect.~\ref{sres}, and present our interpretation
in Sect.~\ref{sdis}. Finally, our conclusions are drawn in Sect.~\ref{scon}.

\section{Archival data analysis}
\label{sarc}

\begin{figure*}
\centering
\resizebox{18cm}{!}{\includegraphics[angle=0]{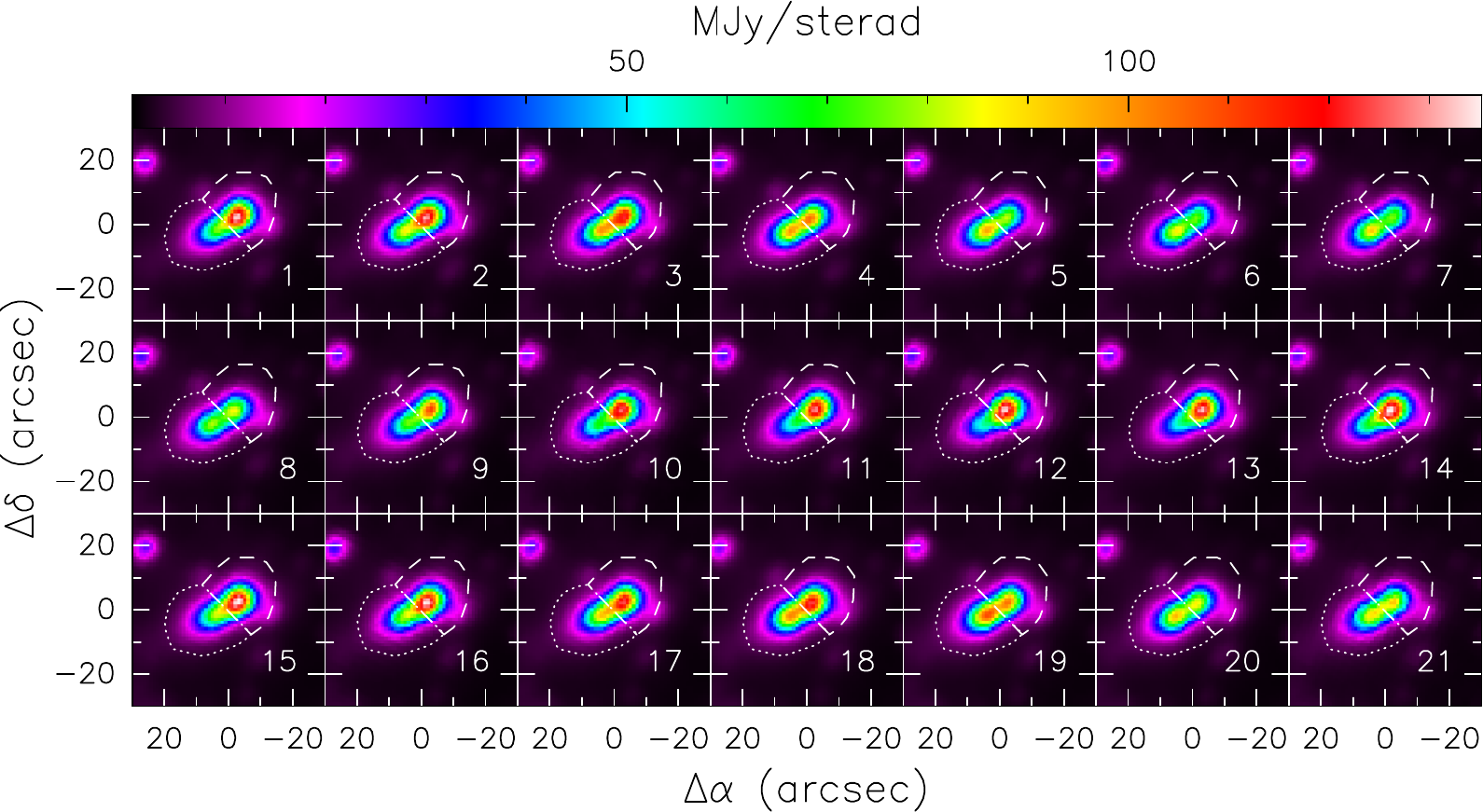}}
\caption{
NEOWISE images of the 3.4~\mic\ emission towards \I\ at 21 epochs. The
number in the bottom right corner of each panel indicates the observation date
as given in the first column of Table~\ref{tneo}.
The offsets are relative to position
$\alpha$(J2000)=$20^{\rm h}14^{\rm m}26\fs0364$,
$\delta$(J2000)=41\degr13\arcmin32\farcs516.
The dashed and dotted patterns outline the polygons used to derive the
mid-IR flux densities of, respectively, the SE (red-shifted) lobe and NW
(blue-shifted) lobe of the bipolar outflow.
}
\label{fneo}
\end{figure*}

The NEOWISE\footnote{https://irsa.ipac.caltech.edu/Missions/wise.html}
(Mainzer et al.~\cite{main11,main14}) database was exploited to
obtain multi-epoch images of the \I\ region. In particular, we used
the single-exposure high-quality images ({\em qual\_frame} = 5 or
10) at band W1 (3.4~\mic) containing the position of the source,
spanning a period from 2014 to 2024. We ignored the W2 band at
4.6~\mic, because the emission is partially saturated.  The images
acquired within a few days were averaged together with the ICORE
tool\footnote{https://irsa.ipac.caltech.edu/applications/ICORE/} to improve
the signal-to-noise ratio. In this way we obtained 21 images at regular intervals
of $\sim$6~months. The corresponding dates are listed in columns 2 and
3 of Table~\ref{tneo}, while the images are shown in Fig.~\ref{fneo}. In
the same table, we also give the positions of the emission peaks obtained
with a 2-D Gaussian fit and the flux densities measured inside the two
polygons drawn in each panel of Fig.~\ref{fneo}. These polygons encompass
the emission from the NW, blue-shifted (flux $S_\nu^{\rm B}$), and SE, red-shifted
(flux $S_\nu^{\rm R}$), lobes of the outflow from \I.

\begin{table}
\caption{Peak positions and flux densities obtained from the NEOWISE data (unless
         otherwise specified).}
\label{tneo}
\tiny
\begin{tabular}{ccccccc}
\hline
\hline
\#$^a$ & date       & MJD   &  $\Delta\alpha$$^b$   &  $\Delta\delta$$^b$   & $S_\nu^{\rm B}$ & $S_\nu^{\rm R}$ \\
 & {\tiny (yy/mm/dd)} & (days) & (arcsec)       &  (arcsec)         & (mJy)        & (mJy) \\
\hline
\dots & 05/06/13$^c$& 53534 &  0.34$\pm$0.26 &  0.86$\pm$0.19 & 408$\pm$20 & 367$\pm$18 \\
\dots & 10/05/15$^d$& 55331 &  3.01$\pm$0.33 &--0.74$\pm$0.22 & 143$\pm$6 & 221$\pm$9 \\
\dots & 10/11/10$^d$& 55510 &  2.35$\pm$0.46 &--0.47$\pm$0.30 & 161$\pm$8 & 210$\pm$12 \\
 1 & 14/05/16 & 56793 &--0.90$\pm$0.28 &  1.38$\pm$0.19 & 245$\pm$12 & 169$\pm$7 \\
 2 & 14/11/13 & 56974 &  0.52$\pm$0.28 &  0.79$\pm$0.19 & 232$\pm$13 & 205$\pm$11 \\
 3 & 15/05/16 & 57158 &  0.37$\pm$0.27 &  0.59$\pm$0.18 & 227$\pm$12 & 210$\pm$10 \\
 4 & 15/11/08 & 57334 &  1.85$\pm$0.29 &--0.09$\pm$0.19 & 181$\pm$10 & 217$\pm$12 \\
 5 & 16/05/14 & 57522 &  2.07$\pm$0.32 &--0.42$\pm$0.21 & 157$\pm$7 & 208$\pm$10 \\
 6 & 16/11/01 & 57693 &  2.73$\pm$0.37 &--0.59$\pm$0.24 & 138$\pm$7 & 197$\pm$11 \\
 7 & 17/05/13 & 57886 &  1.94$\pm$0.35 &--0.36$\pm$0.23 & 152$\pm$7 & 194$\pm$8 \\
 8 & 17/10/29 & 58055 &  1.77$\pm$0.39 &--0.02$\pm$0.25 & 160$\pm$7 & 185$\pm$8 \\
 9 & 18/05/15 & 58253 &--0.12$\pm$0.34 &  0.88$\pm$0.22 & 204$\pm$10 & 170$\pm$7 \\
 10 & 18/10/26 & 58417 &--0.23$\pm$0.31 &  1.09$\pm$0.20 & 222$\pm$17 & 171$\pm$11 \\
 11 & 19/05/14 & 58617 &--1.34$\pm$0.30 &  1.61$\pm$0.21 & 236$\pm$13 & 149$\pm$7 \\
 12 & 19/10/25 & 58781 &--0.87$\pm$0.27 &  1.53$\pm$0.18 & 237$\pm$14 & 155$\pm$8 \\
 13 & 20/05/14 & 58983 &--1.54$\pm$0.28 &  1.70$\pm$0.19 & 241$\pm$10 & 144$\pm$5 \\
 14 & 20/10/24 & 59146 &--0.53$\pm$0.26 &  1.30$\pm$0.17 & 244$\pm$13 & 172$\pm$9 \\
 15 & 21/05/14 & 59348 &--0.67$\pm$0.28 &  1.23$\pm$0.18 & 254$\pm$15 & 185$\pm$9 \\
 16 & 21/10/25 & 59512 &  0.22$\pm$0.28 &  0.89$\pm$0.19 & 233$\pm$15 & 196$\pm$12 \\
 17 & 22/05/14 & 59713 &  0.07$\pm$0.29 &  0.75$\pm$0.19 & 231$\pm$12 & 201$\pm$10 \\
 18 & 22/10/24 & 59876 &  1.28$\pm$0.27 &  0.25$\pm$0.18 & 219$\pm$12 & 233$\pm$12 \\
 19 & 23/05/15 & 60079 &  1.38$\pm$0.28 &--0.04$\pm$0.18 & 199$\pm$9 & 229$\pm$10 \\
 20 & 23/10/25 & 60242 &  2.50$\pm$0.33 &--0.40$\pm$0.21 & 157$\pm$7 & 211$\pm$9 \\
 21 & 24/05/14 & 60444 &  1.33$\pm$0.34 &  0.04$\pm$0.21 & 178$\pm$13 & 200$\pm$14 \\
\hline
\end{tabular}

\vspace*{1mm}
$^a$~epoch number as indicated in Fig.~\ref{fneo} \\
$^b$~offsets with respect to reference position $\alpha$(J2000)=$20^{\rm h}14^{\rm m}26\fs0364$, $\delta$(J2000)=41\degr13\arcmin32\farcs516 \\
$^c$~Spitzer/IRAC data at 3.6~\mic\ from Qiu et al.~(\cite{qiu08})\\
$^d$~ALLWISE data
\end{table}

To extend the monitoring of the region over a longer time interval,
we applied the same procedure to the ALLWISE data (Wright et
al.~\cite{wri10}) at 3.4~\mic\ acquired in 2010, thus obtaining two
additional images. Moreover, we also retrieved the Spitzer IRAC image at
3.6~\mic\ (Werner et al.~\cite{wer04}; Fazio et al.~\cite{faz04}) obtained
by Qiu et al.~(\cite{qiu08}). The corresponding observation dates, positions
and flux densities are given in the first three rows of Table~\ref{tneo}. We
stress that the Spitzer data were smoothed to the same resolution as the
ALLWISE and NEOWISE data before deriving the relevant parameters.

\section{Results}
\label{sres}

\begin{figure}
\centering
\resizebox{8.0cm}{!}{\includegraphics[angle=0]{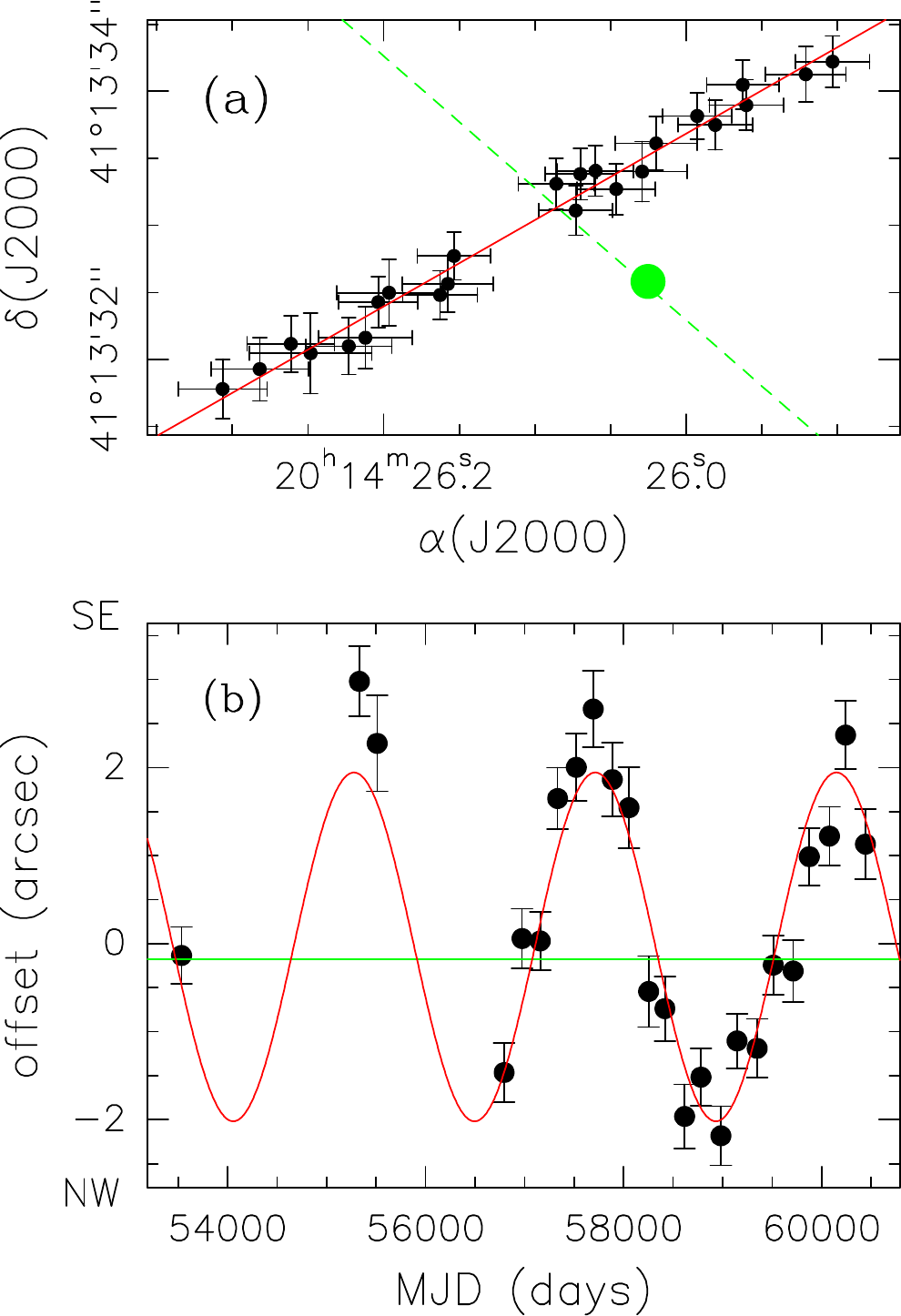}}
\caption{
Result of the 2-D Gaussian fit to the mid-IR emission from \I.
(a) Distribution of the peaks of the emission at the different
epochs (from Table~\ref{tneo}). The full red line with PA=--60\degr\
is a linear fit to the data. The green solid circle marks the peak
of the 1.4~mm continuum, a good proxy for the position of
the protostar. The dashed green line indicates the direction of the plane
of the circumstellar disk.
(b) Offsets of the mid-IR peaks, measured along the red line in the top panel,
as a function of time.
The horizontal green line indicates the offset corresponding to the projection
of the green circle onto the red line in the top panel. The red sinusoid is
the best fit to the data.
}
\label{foffs}
\end{figure}

As already mentioned, Szymczak et al.~(\cite{szy24}) were the first to
detect a periodic variation in \I. Besides discovering the variability of
the \METH\ masers, they also revealed a correlation between the maser emission
and the mid-IR flux from the NEOWISE catalogue. The same authors found
that the peaks of the NEOWISE images are distributed along a straight line
parallel to the outflow axis and the peak position changes
with time. Our results fully confirm this scenario, as illustrated in
Fig.~\ref{foffs}. The linear fit to the peaks has a position angle of
$\sim-60$\degr, consistent with that of the outflow (see e.g. Cesaroni et
al.~\cite{cesa25}) and almost perpendicular to the projected major
axis of the disk, whose position angle is $\sim$49\degr, as shown in
Fig.~\ref{foffs}a. From the same figure one can see that the 1.4~mm continuum
peak is offset from the outflow axis by $\sim$0\farcs9 or $\sim$1500~au.
The (sub)mm continuum peak is a reasonable indication of the position of
the protostar and such an offset suggests that the star might be drifting
from NE to SW, as suggested by Massi et al.~(\cite{massi23}).

In Fig.~\ref{foffs}b we plot the peak position of the mid-IR emission
as a function of time. Clearly, the peak is oscillating between two
extremes and we have fitted the pattern with a sinusoidal curve using
command {\em MFIT} of the GILDAS software\footnote{The GILDAS software
has been developed at IRAM and Observatoire de Grenoble and is available
at http://iram.fr/IRAMFR/GILDAS/}. The fitted function is
$s = \Delta s \, \sin[(2\pi/T)\,t+\Phi]+s_0$,
where $s$ is the offset and $t$ the time, while $\Delta s$, $T$, $\Phi$,
and $s_0$ are the free parameters of the fit. The best fit is obtained
for a period $T$=6.7$\pm$0.2~yr, consistent within the errors with the
period of 6.9$\pm$0.2~yr derived by Szymczak et al.~(\cite{szy24}) for
the \METH\ maser. In the following we assume $T$=6.8~yr, a reasonable
compromise between the two estimates.

The explanation for the periodic pattern in Fig.~\ref{foffs}b can be found
by looking at Fig.~\ref{fneo}. The intensities of the SE (red-shifted)
and NW (blue-shifted) lobes appear to vary but not at the same time:
when one lobe becomes brighter, the other gets dimmer. As a consequence,
when fitting a 2-D Gaussian to the image the peak of the Gaussian tends
to move alternatively from one lobe to the other along the straight line
in Fig.~\ref{foffs}a.

\begin{figure}
\centering
\resizebox{8.0cm}{!}{\includegraphics[angle=0]{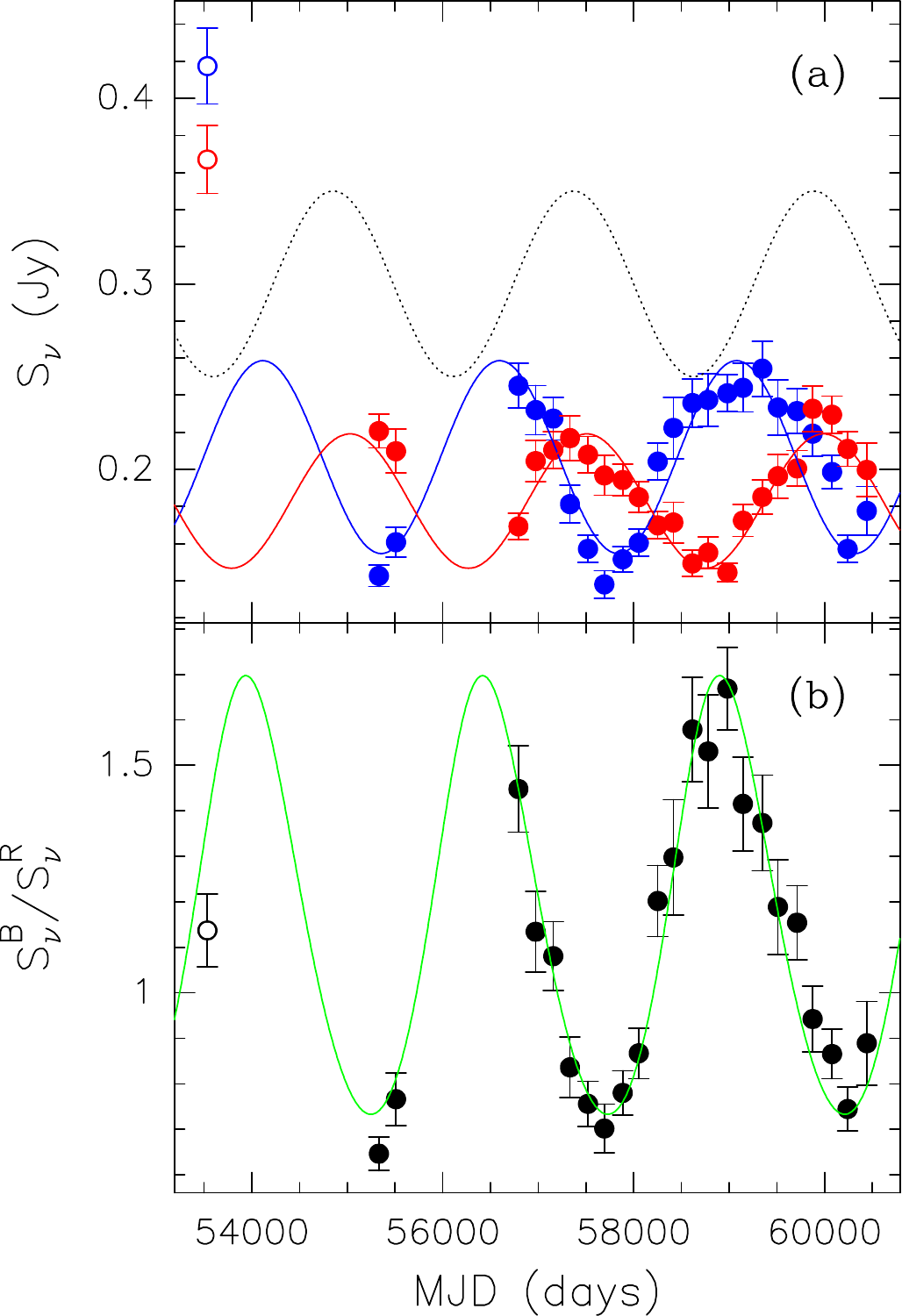}}
\caption{
Variation of the mid-IR emission from the lobes of the outflow.
(a) Integrated flux density over the SE (red points) and NW (blue
points) lobes as a function of time (from Table~\ref{tneo}). The sinusoids
are the best fits to the solid points with the corresponding colour. The empty
points correspond to the Spitzer data and have not been taken into account
in the sinusoidal fits. The dotted
sinusoid is the fit (with arbitrary intensity) to the \METH\ maser intensity obtained from Fig.~1 of
Szymczak et al.~(\cite{szy24}).
(b) Ratio between the fluxes of the NW and SE lobes as a function of
time. The green curve is the ratio between the blue and red sinusoids
in the top panel.
}
\label{fbr}
\end{figure}

This anticorrelated variation is clearly seen in Fig.~\ref{fbr}a, where we
plot the fluxes of the two lobes as a function of time, and is emphasized in
Fig.~\ref{fbr}b, which shows the ratio between the two fluxes. A striking
result that is evident from this plot is that the intensity of the Spitzer
data (first epoch) is by far greater than that of the ALLWISE and NEOWISE
data. We believe that this is not due to the relative calibration of the
two instruments, because for the field stars the flux difference between
the ALLWISE and the Spitzer images is $\sim$2\%, as opposed to $\sim$100\%
for the lobes of the outflow from \I. Such a ratio hints at an episodic
outburst from the massive protostar, possibly due to an accretion event
analogous to that observed, e.g., in S255~NIRS3 (see Caratti o Garatti
et al.~\cite{cagana}).

After excluding the Spitzer data, we fitted two sinusoidal functions
to the fluxes of the two lobes, as already done for the offsets, but
this time the period was fixed to the assumed value of 6.8~yr. The fits
give a phase difference between the two light curves of 2.5$\pm$0.2~yr.
Although a perfect anticorrelation would correspond to a phase difference
of $T/2$=3.4~yr, we can conclude that the emissions from the two lobes are
significantly out of phase.

It is worth noting that, despite the remarkable difference in intensity
between the Spitzer and WISE data, the ratio between the Spitzer fluxes
of the two lobes is consistent with the extrapolation of the fit to the
WISE data only (see Fig.~\ref{fbr}b). This suggests that the observed
variability of the mid-IR emission from \I\ is due to the overlap of
two phenomena: a regular, periodic variation plus a random increase
of intensity that might be related to accretion episodes.

An anticorrelated variation was found also by Szymczak et al.~(\cite{szy24})
between different \METH\ maser features. In particular, they divide the
maser spots into two groups, one to the SE and the other to the NW (see
their Fig. 8). The former they find to be correlated with the near-IR
emission of the SE lobe observed by Massi et al.~(\cite{massi23}) and
we confirm this result as one can see in Fig.~\ref{fbr}a, where the
dotted curve representing the intensity of the \METH\ masers (scaled by
an arbitrary amount to allow comparison with the IR data) closely
follows the best fit (red curve) to the 3.4~\mic\ flux of the SE lobe.

\begin{figure}
\centering
\resizebox{8.0cm}{!}{\includegraphics[angle=0]{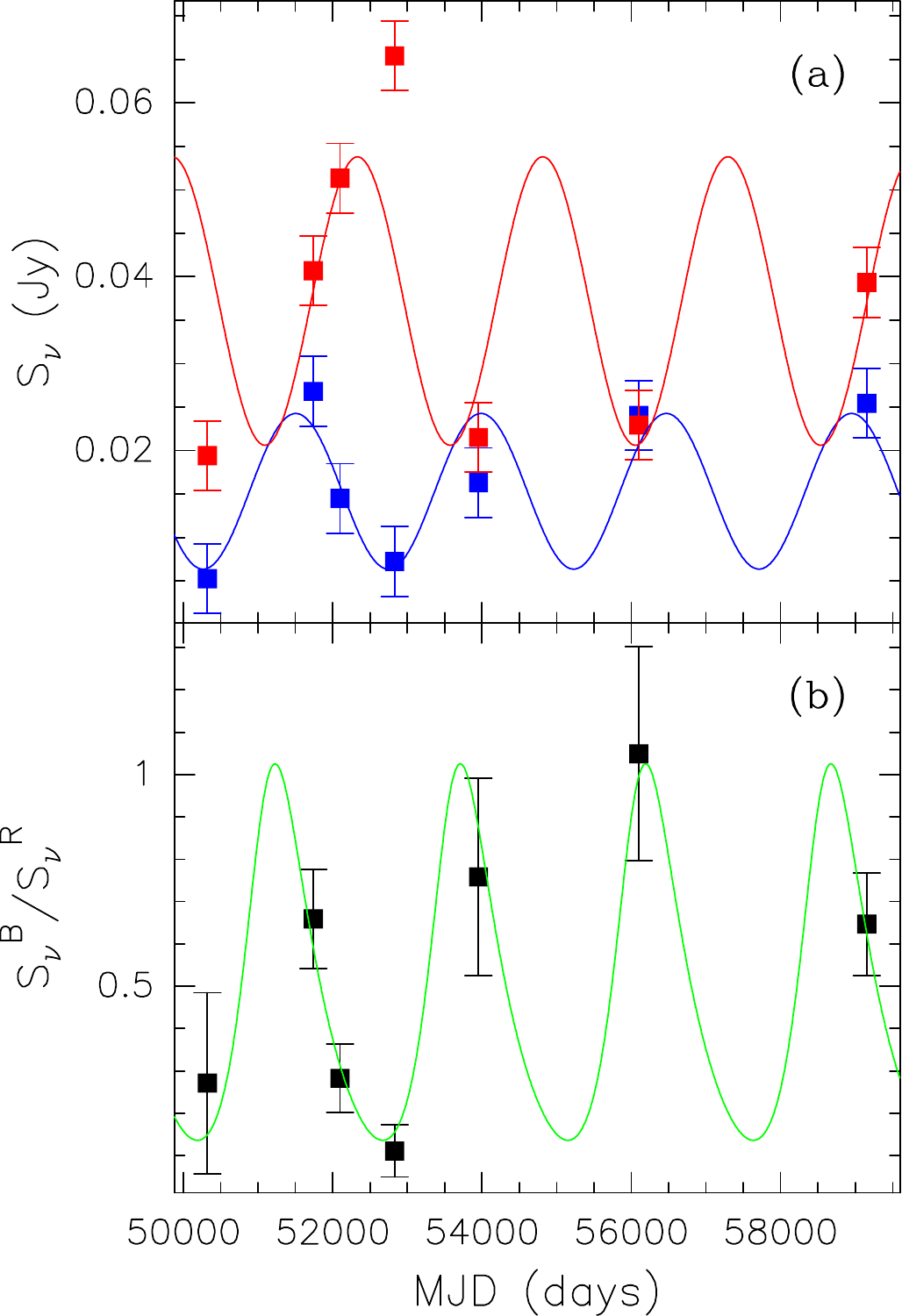}}
\caption{
Same as Fig.~\ref{fbr} for the near-IR data taken from Cesaroni et
al.~(\cite{cesa97}) and Massi et al.~(\cite{massi23}).
}
\label{fnir}
\end{figure}

We also re-analyse the near-IR data of Massi et al.~(\cite{massi23}) in
the light of the NEOWISE results. These authors recognised the existence
of an anticorrelation between the 2.2~\mic\ continuum emission from the
two lobes and hypothesised a regular variability with a period of 12~yr.
In Fig.~\ref{fnir} we show the flux densities obtained by integrating
the near-IR emission inside polygon S1+S2, for the SE lobe, and polygons
N1+N2+N3, for the NW lobe (see Fig.~2 of Massi et al.~\cite{massi23} for
the definition of these polygons). To extend the time interval as much
as possible, we have also added the flux densities obtained from the
K-band image of Cesaroni et al.~(\cite{cesa97}) acquired on August 26, 1996
(MJD=50321~days). As done for NEOWISE, we have fitted the data with two
sinusoids, assuming a fixed period of 6.8~yr. Also in this case
the fits to the intensities of the two lobes and their ratio are satisfactory
thus proving that the anticorrelated variation is seen also in the near-IR.

\begin{figure}
\centering
\resizebox{8.0cm}{!}{\includegraphics[angle=0]{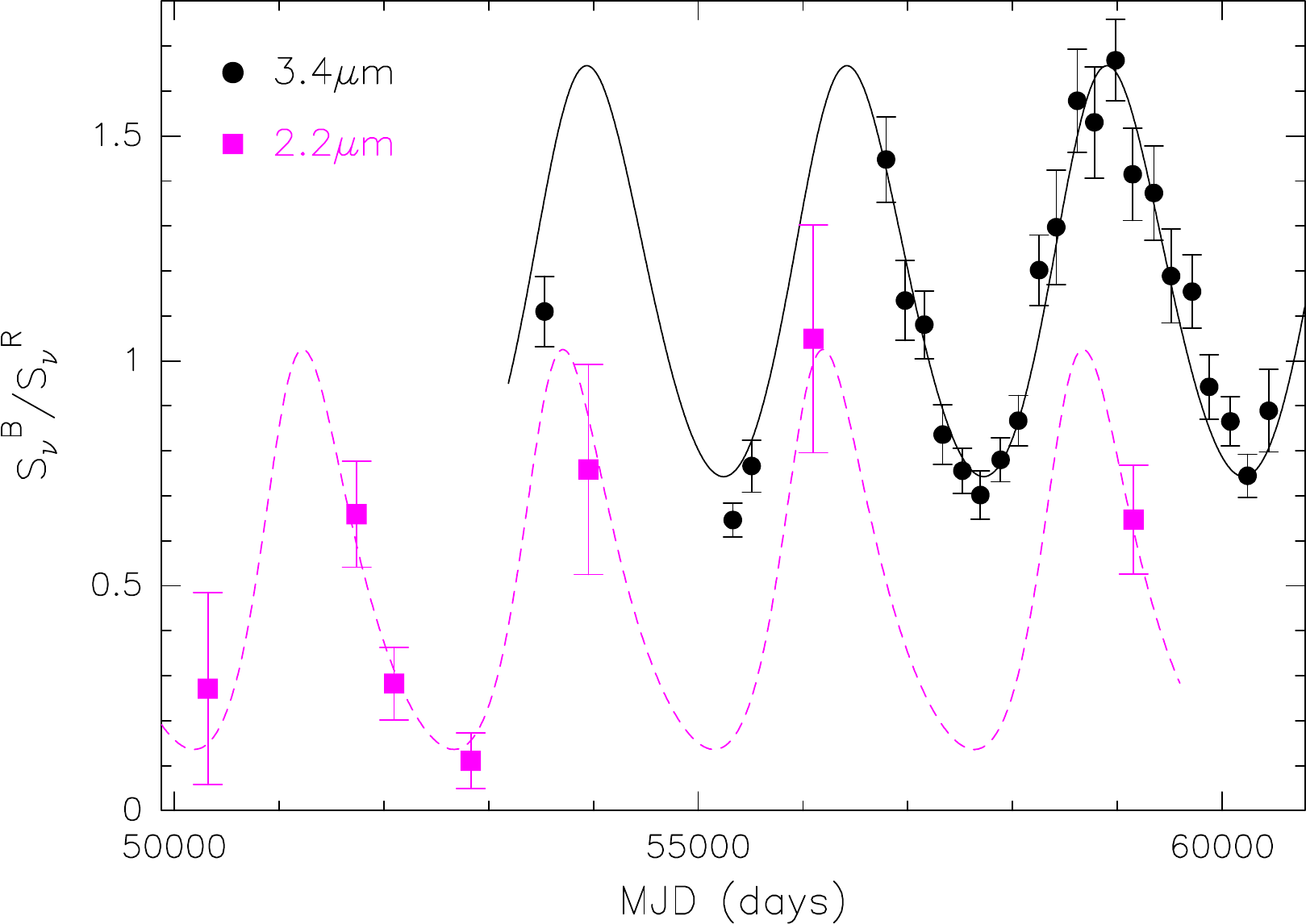}}
\caption{
Comparison between the flux density ratios between the NW and SE lobes
at mid-IR (black circles) and near-IR (magenta squares) as a function
of time. The solid black and dashed magenta curves are the best fits to
the points with the same colour. The data and the fits are the same as
in Figs.~\ref{fbr}b and~\ref{fnir}b.
}
\label{fcomp}
\end{figure}

To lend further support to this conclusion, in Fig.~\ref{fcomp} we compare
the flux ratio between the two lobes in the mid-IR with that in the
near-IR. The phase difference between the two sinusoidal fits is marginal
(0.6$\pm$0.3~yr) and might be due to the poor sampling of the near-IR
light curve.

All in all, the results based on a 28-yr long monitoring confirm the
existence of a mechanism that causes the IR emission from \I\ to undergo
regular variations with a period of $\sim$6.8~yr. We note that Szymczak et
al.~(\cite{szy24})
recognise the existence of a significant variability of the mid-IR emission
from \I, but in the 3.4~\mic\ and 4.6~\mic\ light curves of their Fig.~6 no
clear periodicity is seen. This is due to using the data from the NEOWISE
catalogue, which include the total emission from both lobes. Since the
emissions from the lobes are anticorrelated, using their sum makes it
difficult to identify a periodic pattern. It is also worth noting that
our analysis extends on a longer time interval as it includes also the
last three epochs (in 2023--2024) of the NEOWISE monitoring as well as
two previous measurements by ALLWISE and Spitzer.

\section{Discussion}
\label{sdis}

Our findings imply that
any model to explain the observed variability from \I\ must satisfy
four constraints:
\begin{enumerate}
\item The emission from the outflow lobes must be periodical.
\item The emissions from the two lobes must be anticorrelated.
\item The difference between the peak and the dip of the light curves
 must be $\sim$40\% of the peak (see Fig.~\ref{fbr}a). We define this
 as the ``dimming factor'' $\eta$=0.4.
\item The shape of the light curves must be sinusoidal, in the sense that the
 width of the peaks must be similar to the width of the dips.
\end{enumerate}
With this in mind, in the following we discuss four possible mechanisms that
might explain the observed variability:
pulsations, precession, rotation, and revolution.

\subsection{Pulsations}

Here, with ``stellar pulsations'' we refer generically to any periodic variability
of the luminosity of the star. This can be due to instabilities
in the stellar internal structure or regular accretion events from
the circumstellar disk. Whatever the reason for the periodicity of the
stellar luminosity, this phenomenon cannot explain the light curves in
Fig.~\ref{fbr}, because it is incompatible with the time shift between the
peaks of the emission from the two lobes. In fact, the lobes of the outflow
would be reached at the same time by the photons emitted by the star. In
this case, the observed delay of $\sim$2.5~yr (see Sect.~\ref{sres})
between the peak of the emission from the blue lobe and that from
the red lobe must be equal to the light travel time between the lobes
along the line of sight. The latter is given by the expression
\begin{equation}
t_{\rm light} = \frac{2L\tan\alpha}{c} = 2.5~{\rm yr}   \label{etl}
\end{equation}
where $L$ is the size of a lobe projected onto the plane of the sky, $\alpha$ is the
angle between the outflow axis and the plane of the sky, and $c$ is the
speed of light.

From Fig.~\ref{fneo} one derives $L<10\arcsec=0.08$~pc, hence from Eq.~(\ref{etl})
we obtain $\alpha>78\degr$. Such a value is by far greater than the estimates
of the inclination angle of the outflow obtained from the SiO(5--4) line
(Cesaroni et al.~\cite{cesa25}), the \HM\ line (Massi et al.~\cite{massi23}),
and the 22~GHz \WAT\ masers (Moscadelli et al.~\cite{mosca05}), which range
from 3\degr\ to 8\degr. We thus believe that stellar pulsations are not
a viable explanation for the observed variability.

Finally, it is also worth noting that the model prediction that bloated
protostars undergo a phase of pulsational instability (Inayoshi et
al.~\cite{inay13}), had previously been tested and found to be in good
agreement with the periodic variability observed in the massive protostars
IRAS\,19520+2759 (Pandey et al.~\cite{pand25}) and G353.273+0.641 (Harajiri
et al.~\cite{hara26}). However, this cannot be the case of \I. In fact,
as already noted by Szymczak et al.~(\cite{szy24}), the same model for
a period of 6.8~yr would imply a stellar luminosity of $10^6$~\Lsun,
two orders of magnitude greater than that of \I\ (see Eq.~(1) of Inayoshi
et al.~\cite{inay13}).

\subsection{Precession}

Massi et al.~(\cite{massi23}) and Szymczak et al.~(\cite{szy24}) propose
a scenario where the observed periodic variability is due to the precession
of the innermost part of the circumstellar disk. However, this mechanism
cannot explain any variability if the disk is axially symmetric, because
in this case only the orientation of the precessing disk would change with time, whereas
its inclination would remain the same and the solid angle shadowed by the disk
would be constant. The model works only if the disk is, e.g., optically thick
on one part and thin on the diametrically opposite part. In this case the shadowing of
the stellar photons occurs alternatively in the two lobes thus explaining
why the variation is anticorrelated.

To describe the effect of a precessing disk on the observed luminosity in
a quantitative way, we assume a simple model where only half disk is opaque
in the mid-IR, the outflow lobes are conical with opening angle $\theta_0$,
and the disk is precessing about the outflow axis with an angle $\psi_0$ between
the disk rotation axis and the outflow axis. As detailed in Appendix~\ref{aprec},
one can express the dimming factor as a function of $\theta_0$ and $\psi_0$
through Eq.~(\ref{eadimm}). Therefore, from $\eta(\theta_0,\psi_0)$=0.4
(see constraint n.~3) we derive $\psi_0$ as a function of $\theta_0$, which
implies that the disk inclination must be $\psi_0\ge72\degr$ for any opening
angle of the lobes (see Fig.~\ref{faeta}).

This result has an important implication on the size of the disk.
During the precession, the disk plane wobbles between $-\psi_0$ and $+\psi_0$
in half a period, therefore the velocity at the maximum radius, $R_0$, of
the precessing disk is equal to
\begin{equation}
 \varv = 4 \frac{R_0 \psi_0}{T}.
\end{equation}
In order the disk to remain coherent, such a velocity cannot exceed the
speed of sound, i.e. $\varv\le c_{\rm s}\simeq 1$~\kms. With the additional
condition $\psi_0\ge72\degr$ one eventually obtains
\begin{equation}
 R_0 = \frac{\varv T}{4\psi_0} \le 0.28~{\rm au}.
\end{equation}
Such a small radius is not acceptable for two reasons. It is less than
the expected dust destruction radius for a 12~\Msun\ star (see Table~3 of
Vaidya et al.~\cite{vai09}), which would make the disk optically thin to
mid-IR radiation. And, most important, the rotation period around a 12~\Msun\
star at this radius is $\sim$16~days, by far shorter than the precession
period of 6.8~yr. This result contradicts the hypothesis that a sector
of the disk remains opaque, since the material would be rehashed azimuthally
on a much shorter time scale than the precession period. In conclusion,
we believe that precession of the circumstellar disk cannot explain the
observed periodicity.

Incidentally, we stress that this conclusion does not exclude that part of
the disk could be precessing on a much longer time scale -- as suggested
by the precession period of the associated outflow ($\sim2\times10^4$~yr;
Cesaroni et al.~\cite{cesa05}). Furthermore, the idea that the disk might not be
axially symmetric (as suggested by Cesaroni et al.~\cite{cesa14}) remains
valid, although insufficient by itself to justify the observed variability.

\subsection{Rotation}
\label{srot}

Similar to the Sun spots, also the protostellar surface might present dimmer
regions that, corotating with the protostar, could produce the observed
periodical variations.
This situation is similar to that of weak-line T~Tauri stars, where the
presence of cold spots covering up to $\sim$45\% of the stellar surface
can explain the observed long-term variability, which appears to persist
for many years (Grankin et al.~\cite{gran08}).  In our case, the problem
with this interpretation is that young early-type
stars are known to have rotation periods of several days, by far
shorter than the 6.8~yr period of the \I\ light curves. For example,
Abt et al.~(\cite{abt02}) find that, typically, B-type stars rotate at 25\% of
the breakup velocity, which for a ZAMS star of 12~\Msun\ with a radius of
$\sim$5~\Rsun\ (Panagia~\cite{pana}) corresponds to a rotation period
of $\sim$1.5~days.

However, protostars undergoing accretion are expected to be bloated
(Hosokawa et al.~\cite{hoso09,hoso10}) with radii up to a few 100~\Rsun.
In this case rotation should be slowed down because of conservation of
angular momentum. Since the momentum of inertia of a sphere is proportional
to $R^2$, we conclude that the radius of the bloated protostar should be
equal to $5\,\Rsun\sqrt{6.8\,{\rm yr}/1.5\,{\rm days}}\simeq 200$~\Rsun,
a value comparable to those predicted by theory for massive protostars
undergoing disk accretion (see Hosokawa et al.~\cite{hoso10}). These models
also predict that bloated protostars have a convective zone close to the
surface, consistent with the existence of spots in the stellar photosphere.

Based on the previous considerations, we propose a model where the protostar
has a spot covering half of the equator and extending by $\pm\Delta\theta$
across it. The rotation axis of the star must be inclined by at least
$\Delta\theta$ with respect to the disk axis, so that during the rotation
the spot gradually moves from one half-space with respect to the disk
plane, to the opposite half-space. This makes the emission dimmer for
half a rotation period, alternatively in the two lobes, thus explaining
the anticorrelated variability of the light curves.

In this model the dimming factor $\eta$ is equal to the ratio between
the solid angle subtended by the spot with respect to the center of the
protostar, $\Omega_{\rm S}=2\pi\sin\Delta\theta$, and the solid angle
of an hemisphere, $2\pi$. Therefore $\eta=\sin\Delta\theta=0.4$ and one
obtains $\Delta\theta\simeq24\degr$.

As demonstrated in Appendix~\ref{arota},
we can express the flux densities of the two lobes with Eq.~(\ref{eaflux}):
\begin{eqnarray}
 S_\nu^{\rm B}(\phi) & = & S_\nu^{\rm B,max} \left(1-\frac{\pi-|\phi(t;\Phi)|}{\pi}\eta\right) \label{eblue} \\
 S_\nu^{\rm R}(\phi) & = & S_\nu^{\rm R,max} \left(1-\frac{\pi-|\phi(t;\Phi+\Delta\Phi)|}{\pi}\eta\right) \label{ered}
\end{eqnarray}
where $S_\nu^{\rm B,max}$ and $S_\nu^{\rm R,max}$ are assumed to be constant
in time, $\Phi$ is the phase of the spot (see Eq.~\ref{eaphip}), and
we allow for a phase difference, $\Delta\Phi$, between the two lobes.

\begin{figure}
\centering
\resizebox{8.0cm}{!}{\includegraphics[angle=0]{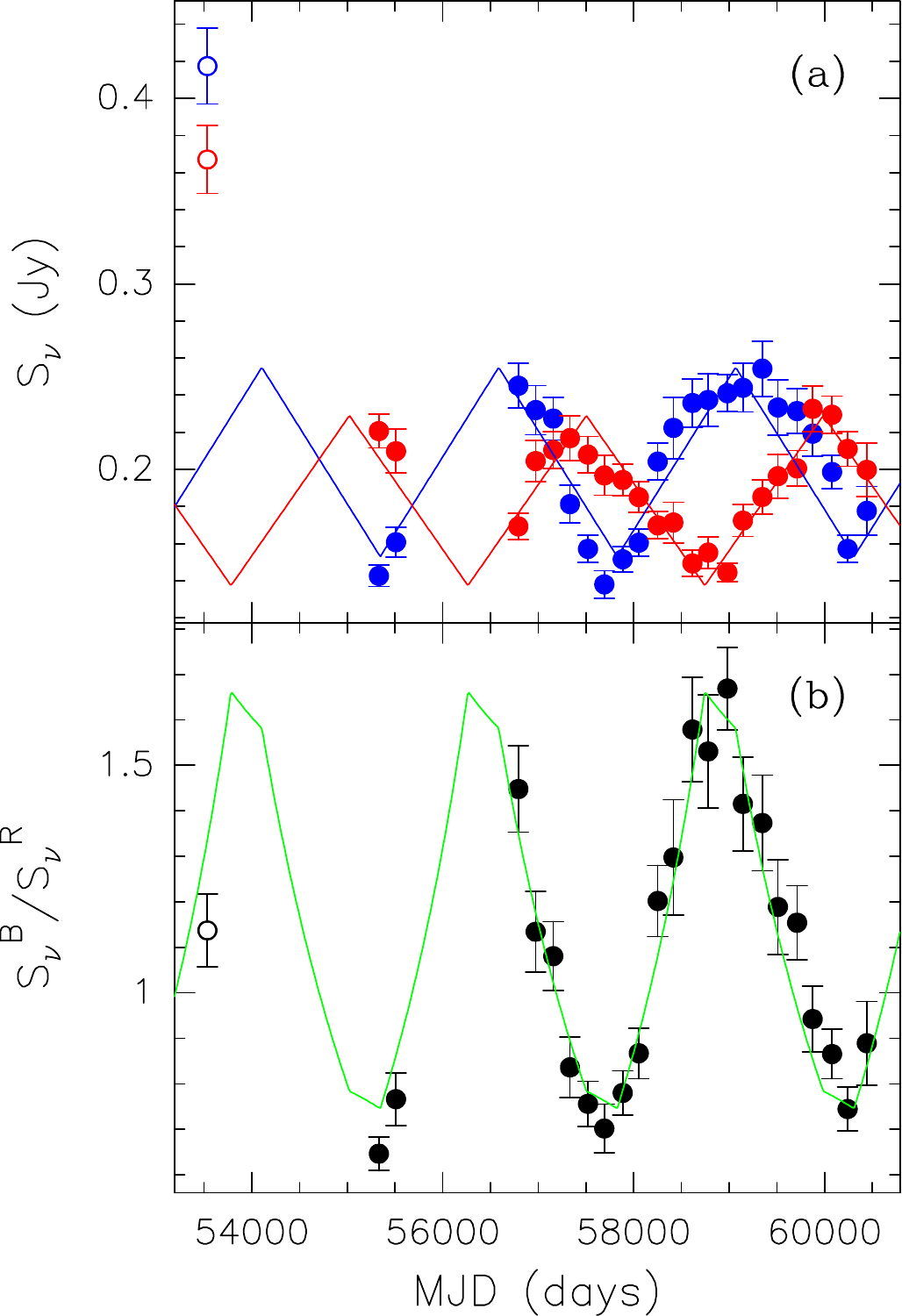}}
\caption{
Same as Fig.~\ref{fbr} with the solid blue, red, and green lines representing
the fit obtained from Eqs.~(\ref{eblue}) and~(\ref{ered}).
}
\label{fspf}
\end{figure}

Using Eqs.~(\ref{eblue}) and~(\ref{ered}), we have fitted the fluxes of
the two outflow lobes simultaneously. The free parameters of the fit are
$S_\nu^{\rm B}$ and $S_\nu^{\rm R}$, $\Phi$, and $\Delta\Phi$, while we
fix the period $T$=6.8~yr. The best fit to the light curves, obtained
for $S_\nu^{\rm B}$=0.255$\pm$0.006~Jy, $S_\nu^{\rm R}$=0.229$\pm$0.005~Jy,
and $\Delta\Phi$=4.0$\pm$0.3~yr, is shown in Fig.~\ref{fspf}, where we
also show the corresponding ratio. Although the fitted curve is clearly very
schematic due to the approximations adopted, its shape is well reproduced.

It is also worth noting that in this model any random variation of the
stellar luminosity (due, e.g., to accretion outbursts) enters in the
terms $S_\nu^{\rm B}$ and $S_\nu^{\rm R}$ as a multiplicative factor that
cancels out in the ratio between the fluxes of the two lobes. Therefore
the curve in Fig.~\ref{fspf}b should represent only the periodic variation
of the emission. This idea is supported by the Spitzer data (empty points)
which are consistent with the model in Fig.~\ref{fspf}b, despite the
large discrepancy shown in Fig.~\ref{fspf}a.

With this in mind, we can elaborate more on Eqs.~(\ref{eblue}) and~(\ref{ered}),
by introducing an additional dependence on time as follows:
\begin{eqnarray}
 S_\nu^{\rm B}(t) & = & S_\nu^\ast(t) f_{\rm B} \left(1-\frac{\pi-|\phi(t;\Phi)|}{\pi}\eta\right) \label{ebt} \\
 S_\nu^{\rm R}(t) & = & S_\nu^\ast(t) f_{\rm R} \left(1-\frac{\pi-|\phi(t;\Phi+\Delta\Phi)|}{\pi}\eta\right) \label{ert}
\end{eqnarray}
where we have indicated with $S_\nu^\ast(t)$ the flux density emitted by the
star and with $f_{\rm B}$ and $f_{\rm R}$ the fractions of the stellar flux
that end up into the two lobes. In our previous fit we
have assumed $S_\nu^\ast$=constant, to derive a mean shape
of the periodic light curve. However, the stellar luminosity is likely to vary
more or less randomly in time, unlike the factors $f_{\rm B}$ and $f_{\rm R}$ which
depend on the structure of the lobes and change on much longer time scales than $T$.
Then, from the ratio between Eq.~(\ref{ebt}) and~(\ref{ert}) we obtain
\begin{equation}
{S_\nu^{\rm B}}' = \frac{f_{\rm B}}{f_{\rm R}} {S_\nu^{\rm R}}'  \label{ecorr}
\end{equation}
where we have defined
\begin{eqnarray}
{S_\nu^{\rm B}}' & = & \frac{S_\nu^{\rm B}}{1-\frac{\pi-|\phi(t;\Phi)|}{\pi}\eta} \label{ebc} \\
{S_\nu^{\rm R}}' & = & \frac{S_\nu^{\rm R}}{1-\frac{\pi-|\phi(t;\Phi+\Delta\Phi)|}{\pi}\eta}. \label{erc}
\end{eqnarray}
In practice, these would be the flux densities of the lobes after removing
the dimming due to the starspot. Equation~(\ref{ecorr}) indicates that
the corrected fluxes must be correlated. This is exactly what one would
expect if there were no spot on the stellar surface, because both lobes
would be illuminated at the same time by the star.

\begin{figure}
\centering
\resizebox{8.0cm}{!}{\includegraphics[angle=0]{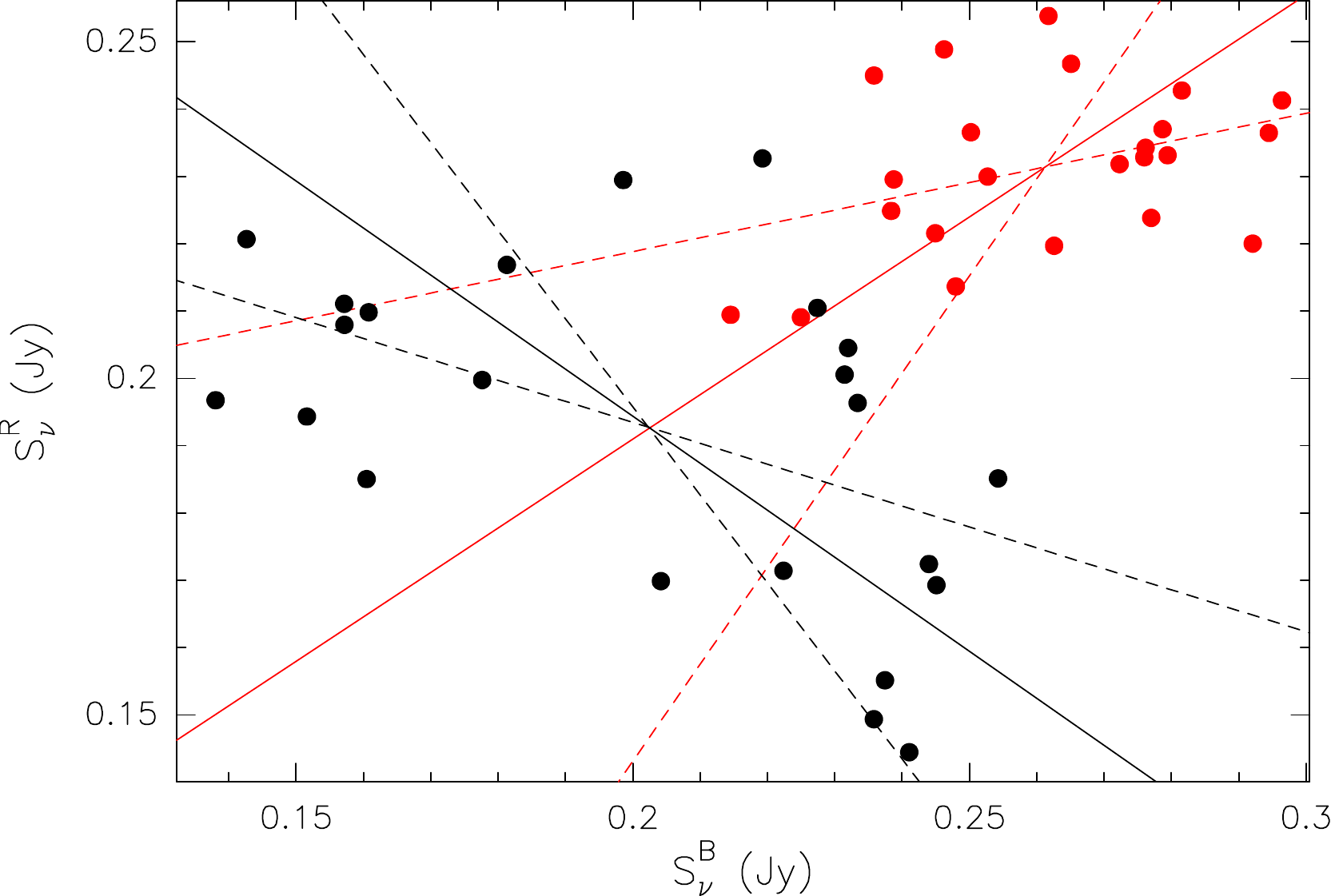}}
\caption{
Flux densities of the red lobe versus those of the blue lobe measured at
23 epochs with ALLWISE and NEOWISE. The Spitzer measurement is not shown.
The black symbols indicate the observed fluxes, $S_\nu^{\rm B}$ and
$S_\nu^{\rm R}$, whereas the red symbols denote the fluxes ${S_\nu^{\rm B}}'$
and ${S_\nu^{\rm R}}'$ corrected for the periodic oscillations using
Eqs.~(\ref{ebc}) and~(\ref{erc}). The solid lines are linear fits to the
points with the same colour and the dashed lines indicate the corresponding
uncertainties.
}
\label{fcorr}
\end{figure}

To test this prediction, we compare in Fig.~\ref{fcorr} the flux density
of the red lobe with that of the blue lobe, before (black points) and after
(red points) the corrections. Although with a large spread, it is clear that
the observed fluxes tend to be anticorrelated, consistent with the analysis
of the light curves in Sect.~\ref{sres}. Vice versa, after the correction
${S_\nu^{\rm B}}'$ and ${S_\nu^{\rm B}}'$ are roughly proportional to each
other, as in Eq.~(\ref{ecorr}). We remark that the Spitzer fluxes have not
been included in this plot because of their large deviation from the mean,
but they would obviously reinforce the correlation between
${S_\nu^{\rm B}}'$ and ${S_\nu^{\rm B}}'$.

\subsection{Revolution}

Exactly the same model fit presented in Sect.~\ref{srot} can be obtained
with a body orbiting around the star and intercepting the stellar photons
emitted in the solid angle described by Eq.~(\ref{eaoms}). Such a body could
be a dense dusty filament distributed along the orbit of radius $\sim$8~au,
which corresponds to a revolution period of 6.8~yr around a 12~\Msun\
star. This orbit must be inclined by at least $\Delta\theta=24\degr$ with
respect to the plane of the disk, as for the model in Sect.~\ref{srot}.
It is worth stressing that any compact body with smaller $\Delta\phi$
would produce too steep a variation of the flux that could not fit the
smooth intensity changes observed between the peaks and the dips of the
light curves (see constraint n.~4).

Although a priori it is not possible to reject the existence of such
a filament, it is not easy to justify the presence of a significant
amount of dust at a distance of only 8~au from a 12~\Msun\ star. In fact,
according to Vaidya et al.~(\cite{vai09}) the dust sublimation radius due to
heating by a circumstellar disk around a 10~\Msun\ star is at least 10~au.
We conclude that periodic shadowing by a body orbiting around the protostar
is not a plausible explanation for the observed flux variations.

\section{Summary and conclusions}
\label{scon}

We have presented robust evidence for a periodic variability of the
mid-IR emission from the massive protostar \I. The flux densities of the
two outflow lobes obtained from Spitzer, ALLWISE, and NEOWISE database
appear to vary over a couple of decades following a sinusoidal light
curve with period of $\sim$6.8~yr. In particular, the emissions from
the two lobes are anticorrelated, with a phase difference of 2.5$\pm$0.2~yr.
These results are confirmed also by the data obtained by Massi et
al.~(\cite{massi23}) and Cesaroni et al.~(\cite{cesa97}) in the near-IR.

We propose four possible interpretations of the results, based on periodic
variability of the stellar luminosity, precession of the circumstellar disk,
rotation of the star with a large spot, and shadowing of the stellar photons
by a body orbiting at 8~au from the protostar. The only acceptable model that
satisfies all the characteristics of the observed light curves is that involving
a dark spot covering 20\% of the whole stellar surface. The 6.8~yr-long
rotation period can be explained with angular momentum conservation if
the protostar has expanded up to a radius of $\sim$200~\Rsun, as predicted
by theory. Therefore, our study provides support for the existence of bloated massive
protostars, suggesting additional methodology for characterizing these
sources, through, e.g., the detection of slow rotation and periodic modulation
of emission due to starspots, presumably associated with a phase of active
surface convection.

We conclude that in all likelihood \I\ is a slowly rotating bloated massive
protostar. In this case the low temperature and faint Lyman continuum of the
protostar are consistent with the lack of a detectable \HII\ region around
\I. While the existence of a large spot on the surface of a such a star may
seem weird, we believe that it remains the most likely explanation for the
observed periodic anticorrelated variability. After all, one should bear in
mind Sherlock Holmes' famous statement that {\it when you have eliminated
the impossible, whatever remains, however improbable, must be the truth}
(The Sign of the Four, 1890).

\begin{acknowledgements}
 I am indebted to Daniele Galli for insightful discussions and valuable
 suggestions which improved the quality of this article, as well as to Rino
 Bandiera for creating Fig.~\ref{fskea} thanks to his unrivalled expertise
 in the {\em Mathematica} software. I also thank Fabrizio Massi for kindly
 providing me with the near-IR flux densities used in this study and Eugenio
 Schisano for useful suggestions on the Spitzer and WISE data.
 This publication makes use of data products from the Wide-field Infrared
 Survey Explorer, which is a joint project of the University of California,
 Los Angeles, and the Jet Propulsion Laboratory/California Institute
 of Technology, and NEOWISE, which is a project of the Jet Propulsion
 Laboratory/California Institute of Technology. WISE and NEOWISE are funded
 by the National Aeronautics and Space Administration.
 This work is also based in part on archival data obtained with the Spitzer
 Space Telescope, which was operated by the Jet Propulsion Laboratory,
 California Institute of Technology under a contract with NASA.
\end{acknowledgements}

\begin{appendix}

\section{Dimming factor due to precessing disk}
\label{aprec}

\begin{figure}
\centering
\resizebox{8.5cm}{!}{\includegraphics[angle=0]{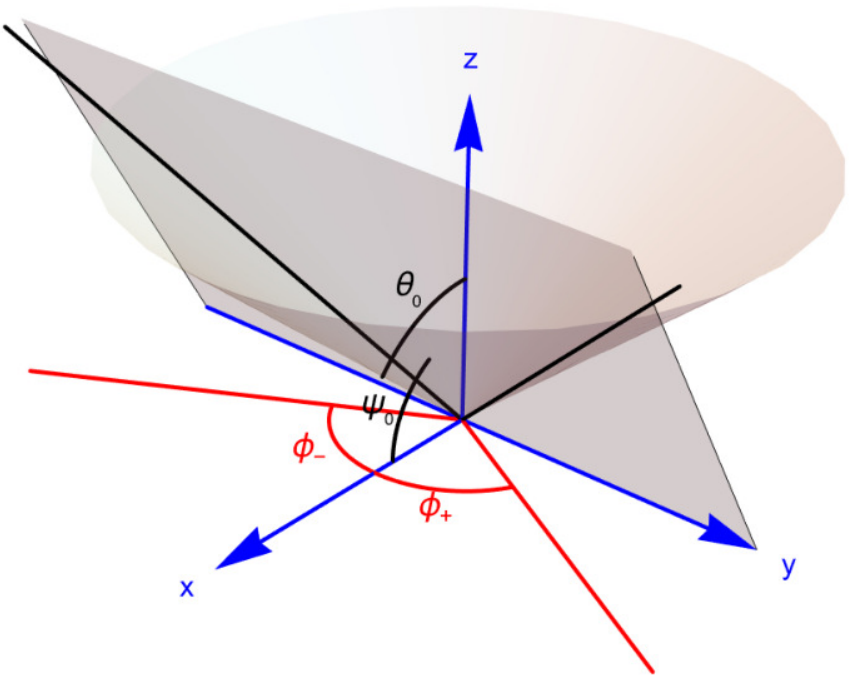}}
\caption{
Sketch of the model consisting of a conical lobe intersected by the plane of
the disk. The axis of the cone coincides with the $z$ axis and the plane
contains the $y$ axis. The black lines are the intersections between
the plane and the cone, while the red lines are their projections onto
the $x,y$ plane. Also indicated are the angle $\psi_0$ between the plane
$x,y$ and the plane of the disk, the opening angle $\theta_0$ of the cone,
and the angles $\phi_\pm$ between the $x$ axis and the red lines.
}
\label{fskea}
\end{figure}

We want to estimate the effect of shadowing of a circumstellar disk on the
stellar photons emitted inside an outflow lobe. In our assumptions the lobe
is a cone with opening angle $0\le\theta_0\le\pi/2$ and vertex coincident
with the star, and the disk axis forms an angle $0\le\psi_0\le\pi/2$ with
the outflow axis (see Fig.~\ref{fskea}). Since the disk can intercept part
of the photons from the star only if it intersects the lobe of the outflow,
we impose the condition $\psi_0>\pi/2-\theta_0$.

The dimming factor, namely the ratio between the flux intercepted by
the disk and the total flux emitted by the star inside the conical lobe,
is equal to the ratio between the solid angle between the disk plane and
the cone, $\Omega$, and the total solid angle of the cone, $\Omega_{\rm
cone}$. In spherical coordinates, the latter is given by the expression
\begin{equation}
 \Omega_{\rm cone} = \int_0^{2\pi} \d\phi \int_0^{\theta_0} \sin\theta \, \d\theta = 2\pi\,(1-\cos\theta_0)
 \label{ealobe}
\end{equation}
while to calculate the former we need to find the intersection between the
cone and the plane. This is obtained from the equations
\begin{eqnarray}
 z & = & \tan\psi_0~x \label{eapl} \\
 z & = & \cot\theta_0 \sqrt{x^2+y^2}
\end{eqnarray}
where we have assumed that the plane of the disk contains the $y$ axis.
We remark that we consider only one of the lobes, for $z>0$, and the intersections
lie in the hemispace $x>0$.
The solution of these equations gives
\begin{equation}
 y = \pm\frac{\sqrt{\tan^2\psi_0-\cot^2\theta_0}}{\cot\theta_0} \, x
\end{equation}
which are the projections of the intersections onto the $x,y$ plane
(red lines in Fig.~\ref{fskea}).
The corresponding azimuthal angles are
\begin{equation}
 \phi_\pm = \pm\arctan\left(\frac{\sqrt{\tan^2\psi_0-\cot^2\theta_0}}{\cot\theta_0}\right).
     \label{eaphi}
\end{equation}

We also need to express the angle $\psi$ between the plane $x,y$ and
a generic line lying on the plane of the disk and passing through the
origin, as a function of the azimuthal angle $\phi$. For a generic point
$(x,y,z)$, the projection of its distance from the origin onto the $x,y$
plane is given by the expression
\begin{equation}
 \sqrt{x^2+y^2} = \cos\psi \sqrt{x^2+y^2+z^2}.   \label{eapsi}
\end{equation}
Substituting $y=x\,\tan\phi$ and Eq.~(\ref{eapl}) into Eq.~(\ref{eapsi}) we obtain
\begin{equation}
 \cos\psi(\phi) = \sqrt{\frac{1+\tan^2\phi}{1+\tan^2\phi+\tan^2\psi_0}}
\end{equation}
and therefore
\begin{equation}
 \sin\psi(\phi) = \frac{\tan\psi_0}{\sqrt{1+\tan^2\phi+\tan^2\psi_0}}.   \label{easinpsi}
\end{equation}

Now, we can calculate the solid angle between the plane and the cone:
\begin{eqnarray}
 \Omega & = & \int_{\phi_-}^{\phi_+} \d\phi \int_{\frac{\pi}{2}-\psi(\phi)}^{\theta_0} \sin\theta \, \d\theta \nonumber \\
        & = & \int_{\phi_-}^{\phi_+} \left[ \sin\psi(\phi)-\cos\theta_0 \right] \d\phi \nonumber \\
        & = & 2 \tan\psi_0\int_{0}^{\phi_+} \frac{\d\phi}{\sqrt{1+\tan^2\phi+\tan^2\psi_0}} \nonumber \\
	& &   - 2\cos\theta_0\arctan\left( \frac{\sqrt{\tan^2\psi_0-\cot^2\theta_0}}{\cot\theta_0} \right) \nonumber \\
        & = & 2 \tan\psi_0 \frac{\sqrt{\tan^2\psi_0\cos(2\phi_+)+\tan^2\psi_0+2}}{\sqrt{2}\tan\psi_0\cos\phi_+\sqrt{\tan^2\psi_0+\tan^2\phi_++1}} \nonumber \\
	& &   \times  \arctan\left( \frac{\tan\psi_0\sin\phi_+}{\sqrt{1+\tan^2\psi_0\cos^2\phi_+}} \right)  \nonumber \\
	& &   - 2\cos\theta_0\arctan\left( \frac{\sqrt{\tan^2\psi_0-\cot^2\theta_0}}{\cot\theta_0} \right).   \label{eaom}
\end{eqnarray}
Here we have used the expression of $\sin\psi$ in Eq.~(\ref{easinpsi})
and the property $\sin\psi(\phi)=\sin\psi(-\phi)$.

From Eq.~(\ref{eaphi}) one obtains
\begin{eqnarray}
\cos\phi_+ & = & \frac{1}{\sqrt{1+\tan^2\phi_+}} = \frac{\cot\theta_0}{\tan\psi_0} \\
\sin\phi_+ & = & \sqrt{1-\cos^2\phi_+} = \frac{\sqrt{\tan^2\psi_0-\cot^2\theta_0}}{\tan\psi_0}.
\end{eqnarray}
By substituting these and Eq.~(\ref{ealobe}) into Eq.~(\ref{eaom}), after
some algebra one derives the expression for the dimming factor:
\begin{eqnarray}
\eta & = & \frac{\Omega}{\Omega_{\rm cone}} ~=~ \frac{1}{\pi(1-\cos\theta_0)} \left[
       \arctan\left(\sqrt{\frac{\tan^2\psi_0-\cot^2\theta_0}{1+\cot^2\theta_0}}\right) \right. \nonumber \\
 & &   \left. - \cos\theta_0 \arctan\left( \frac{\sqrt{\tan^2\psi_0-\cot^2\theta_0}}{\cot\theta_0} \right) \right].
\label{eadimm}
\end{eqnarray}

\begin{figure}
\centering
\resizebox{8.0cm}{!}{\includegraphics[angle=0]{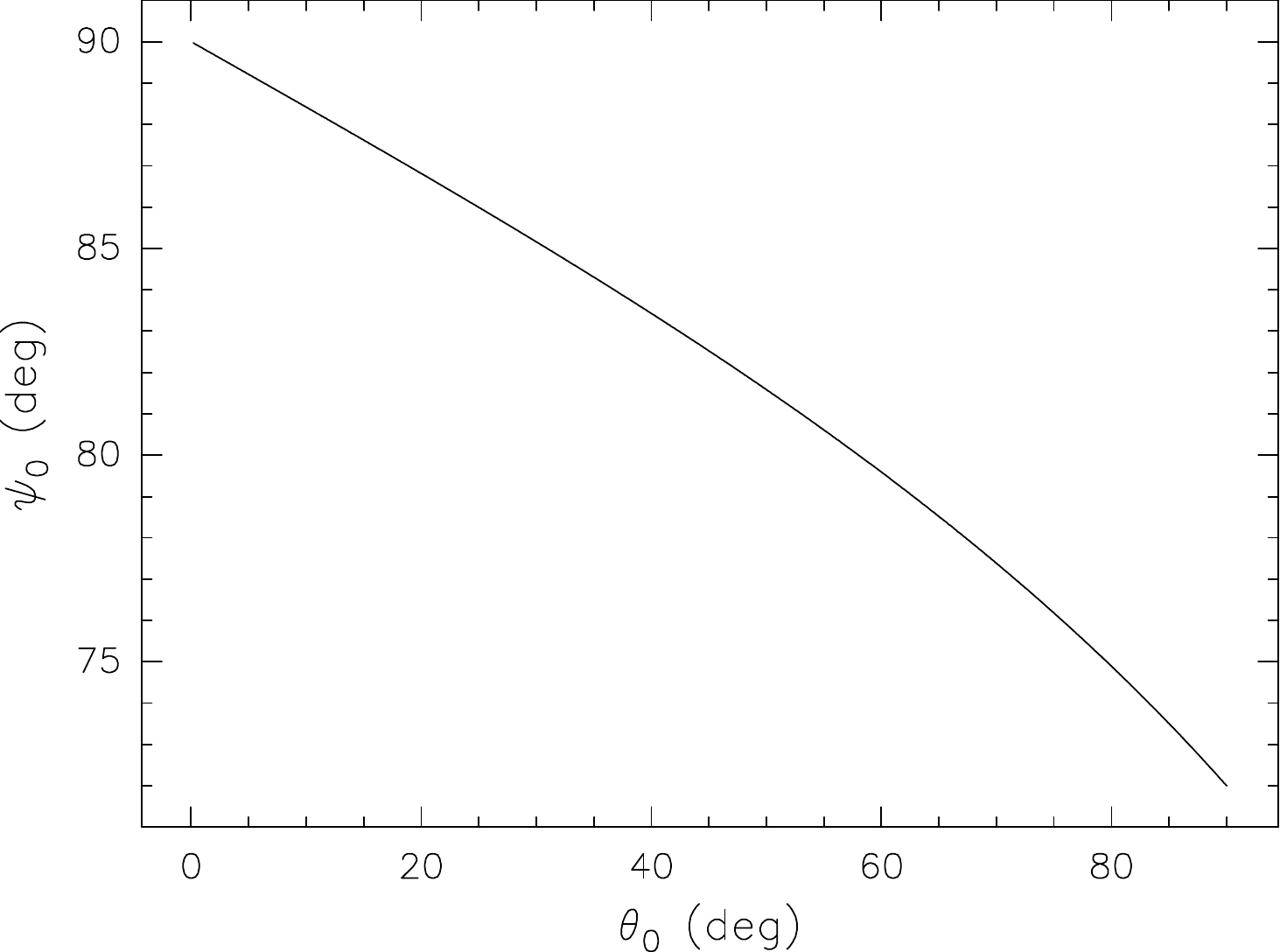}}
\caption{
Inclination angle of the precessing disk as a function of the opening
angle of the outflow lobes, for a dimming factor $\eta=0.4$.
}
\label{faeta}
\end{figure}

We can now relate $\theta_0$ to $\psi_0$ for a given value of $\eta$. In
Fig.~\ref{faeta} we plot $\psi_0$ as a function of $\theta_0$ for $\eta$=0.4,
the value obtained for \I\ (see Sect.~\ref{sdis}). We conclude that in
our case $\psi_0>72\degr$.

\section{Light curve of rotating star with spot}
\label{arota}

We propose a simple model to describe the emission inside one of the
outflow cavities due to a rotating star with a dark spot on its surface. We
assume that the spot extends along half of the equator of the star and
by $\pm\Delta\theta$ in latitude. The rotation axis of the star must
be inclined by at least $\Delta\theta$ with respect to the disk axis,
so that in one rotation period the spot moves from one side with respect
to the disk plane, to the other side.

\begin{figure}
\centering
\resizebox{8.5cm}{!}{\includegraphics[angle=0]{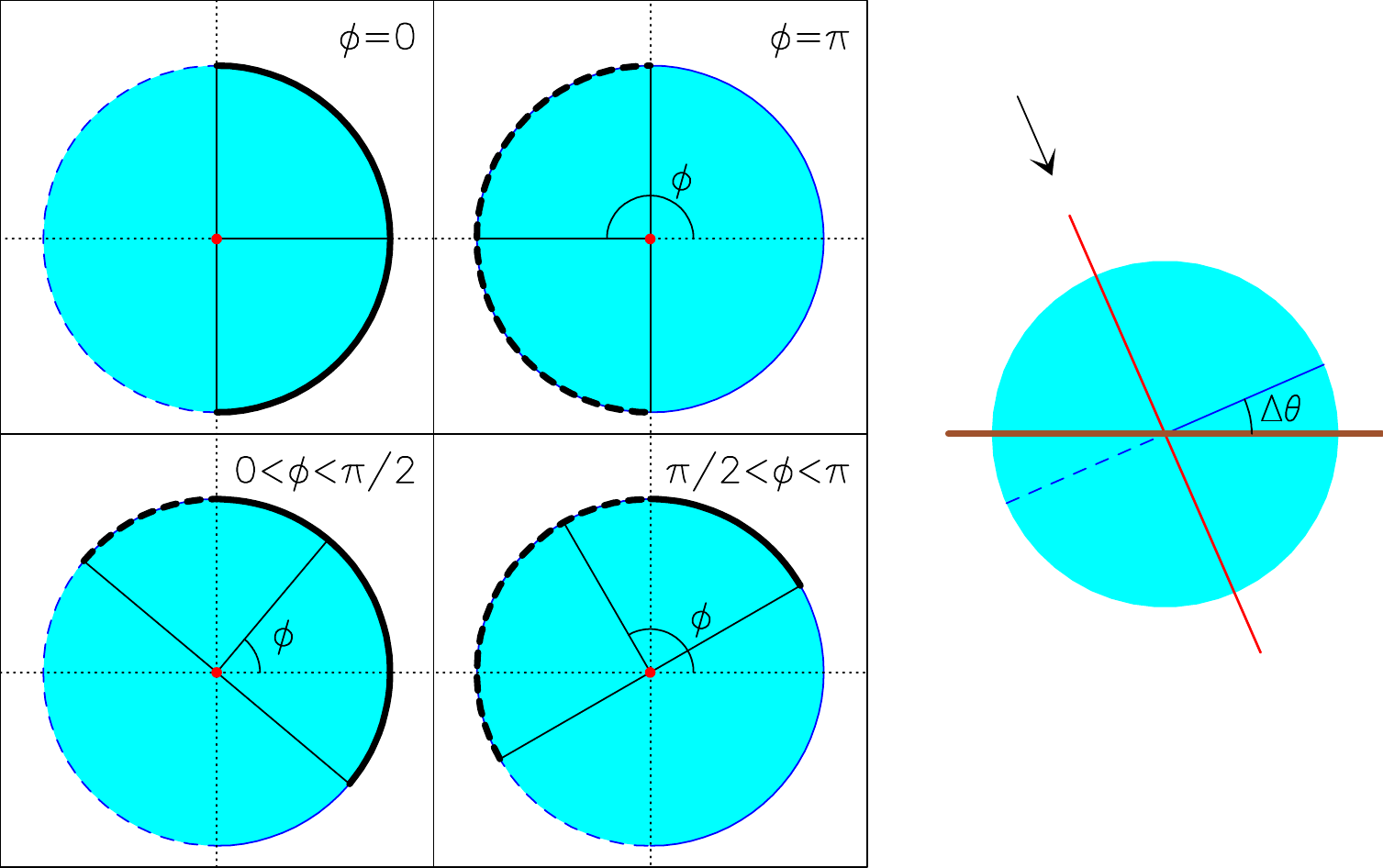}}
\caption{
Sketch of the rotating star with an equatorial spot. The cyan circle is the
star, while the blue and red colours indicate, respectively, the equator
and the rotation axis of the star. The brown line denotes the plane of the
disk seen edge on. Dashed patterns indicate a position under the disk with
respect to the lobe under consideration. The thick black pattern marks the
extension of the spot along the equator, with the solid and dashed arcs
representing, respectively, the part of the spot lying inside the lobe
of interest and that on the opposite side. The pictures inside the boxes
show four template positions of the spot for the azimuthal angles given in
the top right corner of the boxes. The point of view of the four panels is
along the rotation axis, as indicated by the arrow in the figure to the
right, which shows the star as seen by an observer whose line of sight
lies in the plane of the disk and is perpendicular to the star rotation axis.
}
\label{fask}
\end{figure}

The flux emitted in one of the lobes can be expressed as
\begin{equation}
 S_\nu(\phi) = S_\nu^{\rm max} \left( 1 - \frac{\Omega(\phi)}{2\pi} \right)
 \label{easnu}
\end{equation}
where $S_\nu^{\rm max}$ is the maximum flux density emitted when the spot
lies on the opposite side of the disk plane with respect to the lobe, $\Omega$ is the
solid angle subtended by the portion of the spot that lies on the same
side as the lobe, and $\phi$ is the azimuthal angle that we assume to
lie in the range $-\pi<\phi<+\pi$, with $\phi=0$
when the spot lies entirely on the same side as the lobe (i.e. when the minimum
intensity is reached). The solid angle can be computed as
\begin{equation}
\Omega = \Delta\phi \int_{\frac{\pi}{2}-\Delta\theta}^{\frac{\pi}{2}+\Delta\theta} \sin\theta \, \d\theta
       = 2 \,\Delta\phi \, \sin\Delta\theta.
\end{equation}
Here, $\Delta\phi$ is a function of $\phi$ as one can see from the sketch
in Fig.~\ref{fask}, where the $\Delta\phi$ is the angle subtended by the thick black solid arc.
In this fugure, we show four template cases with
different orientations of the spot, for $\phi\ge0$. It is easy to see that in all
cases $\Delta\phi=\pi-\phi$. Analogously, for $\phi<0$ one obtains
$\Delta\phi=\pi+\phi$, so that in general we can write
\begin{equation}
 \Omega(\phi) = 2 (\pi-|\phi|) \sin\Delta\theta.
 \label{eaoms}
\end{equation}

For our purposes it is useful to express $\Delta\theta$ as a function
of the measurable quantity $\eta$, the dimming factor that by definition
is equal to (see constraint n.~3 in Sect.~\ref{sdis})
\begin{equation}
 \eta = \frac{S_\nu^{\rm max}-S_\nu^{\rm min}}{S_\nu^{\rm max}} =
 \frac{S_\nu(\pi)-S_\nu(0)}{S_\nu(\pi)} = \frac{\Omega(0)}{2\pi} = \sin\Delta\theta
 \label{eaeta}
\end{equation}
where $S_\nu^{\rm min}$ is the minimum observed value of the flux. By
substituting Eqs.~(\ref{eaeta}) and~(\ref{eaoms}) into Eq.~(\ref{easnu}),
the expression of the flux takes the form
\begin{equation}
 S_\nu(\phi) = S_\nu^{\rm max} \left( 1 - \frac{\pi-|\phi|}{\pi}\eta \right).  \label{eaflux}
\end{equation}

To fit the data we must express $S_\nu$ as a function of time. We thus define
the angle
\begin{equation}
\phi'(t;\Phi) = 2\pi \, \frac{t}{T} + \Phi   \label{eaphip}
\end{equation}
where $T=6.8$~yr is the rotation period and $\Phi$ is the phase. Finally,
$\phi(t;\Phi)$ is obtained by reducing $\phi'(t;\Phi)$ into the range
between $-\pi$ and $+\pi$.

\end{appendix}


\begin{thebibliography}{}

\bibitem[2007]{abt02}
 Abt, H. A., Levato, H., \& Grosso, M. 2002, ApJ, 573, 359
\bibitem[2008]{caga08}
 Caratti o Garatti, A., Froebrich, D., Eisl\"offel, J., Giannini, T., \& Nisini, B. 2008, A\&A, 485, 137
\bibitem[2017]{cagana}
 Caratti o Garatti, A., Stecklum, B., Garcia Lopez, R., et al. 2017, Nature Physics, 13, 276
\bibitem[1997]{cesa97}
 Cesaroni, R., Felli M., Testi L., Walmsley C.M., \& Olmi L. 1997, A\&A, 325, 725
\bibitem[2005]{cesa05}
 Cesaroni, R., Neri, R., Olmi, L., et al. 2005, A\&A, 434, 1039
\bibitem[2014]{cesa14}
 Cesaroni R., Galli D., Neri R., \& Walmsley C.~M. A\&A, 566, A73
\bibitem[2023]{cesa23}
 Cesaroni, R., Faustini, F., Galli, D., et al. 2023, A\&A, 671, A126
\bibitem[2025]{cesa25}
 Cesaroni, R., Galli, D., Padovani, M., Rivilla, V. M., \& S\'anchez-Monge, \'A. 2025, A\&A, 693, A76
\bibitem[2016]{chen}
 Chen, H.-R. V., Keto, E., Zhang, Q., et al. 2016, ApJ, 823, 125
\bibitem[2004]{faz04}
 Fazio, G. G., Hora, J. L., Allen, L. E., et al. 2004, ApJS, 154, 10
\bibitem[2007]{felli07}
 Felli, M., Brand, J., Cesaroni, R., et al. 2007, A\&A, 476, 373
\bibitem[2015]{fuji15}
 Fujisawa, K., Yonekura, Y., Sugiyama, K., et al. 2015, The Astronomer's Telegram, 8286, 1
\bibitem[2008]{gran08}
 Grankin, K. N., Bouvier, J., Herbst, W., \& Melnikov, S. Yu. 2008, A\&A, 479, 827
\bibitem[2026]{hara26}
 Harajiri, S., Motogi, K., Nakamura, R., et al. 2026, ApJ, 997, 349
\bibitem[2020]{vera}
 Hirota, T., Nagayama, T., Honma, M., et al. 2020, PASJ, 72, 50
\bibitem[2022]{hiro22}
 Hirota, T., Wolak, P., Hunter, T. R., et al. 2022, PASJ, 74, 1234
\bibitem[2009]{hoso09}
 Hosokawa, T. \& Omukai, K. 2009, ApJ, 691, 823
\bibitem[2010]{hoso10}
 Hosokawa, T., Yorke, H. W., \& Omukai, K. 2010, ApJ, 721, 478
\bibitem[2017]{hunt17}
 Hunter, T. R., Brogan, C. L., MacLeod, G., et al. 2017, ApJ, 837, L29
\bibitem[2013]{inay13}
 Inayoshi, K., Sugiyama, K., Hosokawa, T., Motogi, K., \& Yanaka, K. E. I., 2013, ApJ, 769, L20
\bibitem[2011]{johns}
 Johnston, K.G., Keto, E., Robitaille, T. P. Wood, K. 2011, MNRAS, 415, 2953
\bibitem[2026]{kulk26}
 Kulkarni, C. S., Behling, T., Banks, E. E., et al. 2026, ApJ, in press
\bibitem[2024]{lu24}
 Lu, Y., Chen, X., Song, S.-M., et al. 2024, ApJS, 272, 44
\bibitem[2011]{main11}
  Mainzer, A., Bauer, J., Grav, T., et al. 2011, ApJ, 731, 53
\bibitem[2014]{main14}
  Mainzer, A., Bauer, J., Cutri, R. M., et al. 2014, ApJ, 792, 30
\bibitem[2023]{massi23}
 Massi, F., Caratti o Garatti, A., Cesaroni, R., et al. 2023, A\&A, 672, A113
\bibitem[2005]{mosca05}
 Moscadelli, L., Cesaroni, R., \& Rioja, M.J. 2005, A\&A, 438, 889
\bibitem[2011]{mosca11}
 Moscadelli, L., Cesaroni, R., Rioja, M.J., Dodson, R., \& Reid, M.J. 2011, A\&A 526, A66
\bibitem[2025]{neha25}
 Neha, S. \& Sharma, S. 2025, ApJS, 278, 10
\bibitem[1973]{pana}
 Panagia, N. 1973, AJ, 78, 929
\bibitem[2025]{pand25}
 Pandey, R., Palau, A., Serna, J., et al. 2025, MNRAS, 541, 3772
\bibitem[2021]{park21}
 Park, W., Lee, J.-E., Contreras Pe\~na, C., et al. 2021, ApJ, 920, 132
\bibitem[2008]{qiu08}
 Qiu, K., Zhang, Q., Megeath, S.Th., et al. 2008, ApJ, 685, 1005
\bibitem[2024]{szy24}
 Szymczak, M., Durjasz, M., Goedhart, S., et al. 2024, A\&A, 682, A17
\bibitem[2000]{shep}
 Shepherd, D.S., Yu, K.C., Bally, J., \& Testi, L. 2000, ApJ, 535, 833
\bibitem[2016]{steck16}
 Stecklum, B., Caratti o Garatti, A., Cardenas, M. C., et al. 2016, The Astronomer's Telegram, 8732, 1
\bibitem[2019]{sugi19}
 Sugiyama, K., Saito, Y, Yonekura, Y., \& Momose, M. 2019, The Astronomer's Telegram, 12446, 1
\bibitem[2024]{tayo24}
 Tanabe, Y. \& Yonekura, Y. 2024, PASJ, 76, 426
\bibitem[2009]{vai09}
 Vaidya, B., Fendt, C., \& Beuther, H. 2009, ApJ, 702, 567
\bibitem[2004]{wer04}
 Werner, M., Roellig, T., Low, F., et al. 2004, ApJS, 154, 1
\bibitem[2010]{wri10}
 Wright, E.L., Eisenhardt, P.R.M., Mainzer, A.K., et al. 2010, AJ, 140, 1868

\end{thebibliography}
\end{document}